\documentclass[aps, onecolumn, superscriptaddress, 12pt]{revtex4-2}

\usepackage{graphicx}
\usepackage{amsmath}
\usepackage{amssymb}
\usepackage{setspace}
\usepackage{color}
\usepackage{textcomp}
\usepackage{comment}
\usepackage{float}
\usepackage{psfrag}
\usepackage{siunitx}
\usepackage{romannum}
\usepackage{bm}

\begin{document}
\pagenumbering{arabic} 

\title{Compression-thinning behavior of bubble suspensions}

\author{Hu Sun}
\affiliation{School of Astronautics, Beihang University, Beijing 100191, PR China}

\author{Qingfei Fu}
\email{fuqingfei@buaa.edu.cn}
\affiliation{School of Astronautics, Beihang University, Beijing 100191, PR China}
\affiliation{Aircraft and Propulsion Laboratory, Ningbo Institute of Technology, Beihang University, Ningbo 315800, PR China}
\affiliation{National Key Laboratory of Aerospace Liquid Propulsion, Xi’an 710100, PR China}

\author{Chiyu Xie}
\affiliation{School of Astronautics, Beihang University, Beijing 100191, PR China}
\affiliation{Aircraft and Propulsion Laboratory, Ningbo Institute of Technology, Beihang University, Ningbo 315800, PR China}

\author{Bingqiang Ji}
\email{bingqiangji@buaa.edu.cn}
\affiliation{School of Astronautics, Beihang University, Beijing 100191, PR China}
\affiliation{Aircraft and Propulsion Laboratory, Ningbo Institute of Technology, Beihang University, Ningbo 315800, PR China}

\author{Lijun Yang}
\email{yanglijun@buaa.edu.cn}
\affiliation{School of Astronautics, Beihang University, Beijing 100191, PR China}
\affiliation{Aircraft and Propulsion Laboratory, Ningbo Institute of Technology, Beihang University, Ningbo 315800, PR China}
\affiliation{National Key Laboratory of Aerospace Liquid Propulsion, Xi’an 710100, PR China}

\begin{abstract} 
Rheology of bubble suspensions is critical for the prediction and control of bubbly flows in a wide range of industrial processes. It is well-known that the bubble suspension exhibits a shear-thinning behavior due to the bubble shape deformation under pure shear, but how the shear rheology response to dilatation remains unexplored. Here, we report a compression-thinning behavior that the bubble suspension exhibits a decreasing shear viscosity upon compressing. This peculiar rheological behavior is microscopically due to that a shrinking bubble surface effectively weakens the flow resistance of the surrounding liquid. We theoretically propose a constitutive equation for dilute bubble suspensions considering both shear and dilatation effects, and demonstrate that the contribution of dilatation effect on the shear viscosity can be significant at a changing pressure.
\end{abstract}

\maketitle
\section{Introduction} \label{sec:1}
Bubbly flows pervade a wide range of industrial processes in geological, chemical, pharmaceutical, and food engineering \cite{Rodríguez-Rodríguez2015, Dollet2019}. Though the continuous liquid phase is Newtonian, the introduction of discrete bubbles endows the dispersion system with complex non-Newtonian behaviors such as shear-dependent viscosity and elasticity \cite{Stein2002, Llewellin2002, Mader2013, zenit_2018}, producing challenges to the prediction and control of the flow behaviors. Originating with the well-known Einstein equation \cite{einstein_1905, einstein_1911}, rheological models of bubble suspensions have been established with a special focus on the shear viscosity \cite{Taylor1932, Oldroyd1953, Frankel1970, stone1994, Manga1998, Manga2001}, where bubbles are regarded as special incompressible ``inviscid droplets". These models predict the shear-thinning and relaxation behavior of bubble suspensions, which has been confirmed by experiments with simple shear flows \cite{Rust2002a, Lim2004, Joh2010, morini2019, mitrou_2023, ohie2024}. 

However, different from droplets, bubbles are highly compressible. Taylor pointed out that the presence of bubbles makes the suspension compressible and results in a dilatational viscosity \cite{Taylor1954}. Thus, dilatation motion will be triggered in bubbly flows due to the violent pressure variations in practical scenarios such as foam flooding and fuel injection \cite{islam1990, fisher1999, Or2016,bergwerk1959,nurick1976}. In fact, the expansion or compression of bubbles change the shear flow field by introducing an additional centripetal flow and affecting their shape deformation \cite{tanveer_1995, pozrikidis_2001, pozrikidis_2003, crowdy_2003}, and thus may modify the shear rheological properties of the suspension. Unfortunately, due to the incompressible assumption of bubbles, previous models fundamentally ignored the contribution of dilatation motion on the suspension's shear viscosity \cite{Oldroyd1953,Frankel1970}. How the shear rheology of bubble suspensions response to dilatation remains unexplored.

In this work, we theoretically develop a novel constitutive equation for dilute bubble suspensions undergoing a motion of both shear and dilatation. We reported a compression-thinning behavior, a distinct rheological property endowed by bubble compressibility upon a changing pressure, which can significantly influence the suspensions' shear viscosity. Our study highlight the critical role of dilatation motion in mediating the shear rheology of bubble suspensions.

\section{Constitutive Equation}\label{sec:2}
We consider the shear rheology of a fluid element of a dilute suspension with length scales much larger than the bubble size, where the relationship between the bulk stress and the bulk rate-of-strain tensor is established by Batchelor's volume averaging method \cite{Batchelor1970}, which gives
\begin{equation}    \label{eq_integral}
    \left\langle \bm{\tau}  \right\rangle = 2 \mu_l \left\langle \mathbf{e} \right\rangle + \frac{N}{V} \bm{\Sigma} ^g ,
\end{equation} 
where $\left \langle \cdot \right \rangle$ marks averaged parameters in the fluid element, $\bm{\tau}$ is the deviatoric stress tensor, $\mathbf{e}$ is the rate-of-strain tensor, $\mu_l$ is the viscosity of the liquid phase which is Newtonian here. $V$ is the element volume, and $N$ is the bubble number. For simplicity, the bubbles in the fluid element are assumed to be monodisperse without interactions between each other. Based on Gauss's law, the stress contribution of an individual bubble, $\bm{\Sigma} ^g$, is written as \cite{Batchelor1970}
\begin{equation}    \label{eq_Sigma}
    \bm{\Sigma} ^g = \int_S { {\mathbf{n} \cdot \left( {{\bm \sigma}_l \mathbf{r}} \right) - \frac{1}{3}\left( {{\bm \sigma}_l : \mathbf{nr}} \right) \bm{\delta}  - \mu_l \left( {\mathbf{n}{\mathbf u}_l + {\mathbf u}_l \mathbf{n}} \right)} }\:\mathrm{d}S ,
\end{equation}
where the subscripts $l$ and $g$ respectively denote the liquid and gas phases, $\mathbf{n}$ and $\mathbf{r}$ respectively represent the unit outer normal vector and the position vector on the surface $S$, $\bm{\sigma}$ and $\mathbf{u}$ respectively represent the stress tensor and local velocity vector, $\bm \delta$ is the Kronecker tensor. 

To evaluate the influence of bubbles to the macroscopic rheological properties, we consider the microscopic flow of a fluid element containing a small bubble of equivalent initial radius $a$, with a macroscopic velocity $\mathbf{U}$, pressure $P$, shear rate $\dot\gamma$, and rate of change in pressure $\mathrm{d}P/\mathrm{d}t$. In the following, we make all parameters dimensionless using $a$, $\dot \gamma ^{-1}$, and ${\mu _l}\dot \gamma $ as the characteristic length, time, and stress scales, respectively. The liquid flow around the bubble is described by the Stokes equation considering the low Reynolds number \cite{Cox1969},

\begin{equation}    \label{eq_liquid}
    \nabla \cdot \mathbf{u}_l = 0 , \quad
    \nabla p_l = \nabla ^2 \mathbf{u}_l ,
\end{equation}
and the gas in the bubble is assumed to be compressible, inviscid, and ideal, yielding
\begin{equation}    \label{eq_bubble}
    \nabla \cdot \mathbf{u}_g = -\frac{1}{p_g} \frac{\mathrm{d} p_g}{\mathrm{d} t} , \quad
    \nabla p_g = 0,
\end{equation}
where $\nabla$ is the Hamiltonian operator and $p$ is the local pressure. 

We describe the velocities, stresses, and the bubble deformations explicitly through spherical harmonic analysis. The solutions to Stokes Equations [Eq. (\ref{eq_liquid})] were given by Lamb \cite{Lamb1932}, and the velocity inside the bubble is derived as
\begin{equation}    \label{eq_bubbbleVelocity}
    {\mathbf{u}_g} = - \frac{1}{{6{p_g}}}\frac{{\mathrm{d}{p_g}}}{{\mathrm{d}t}} \nabla {r^2} + \sum\limits_{n = 0}^{ + \infty } {\nabla \left( r^n \mathcal{Y}^g_n \right) } ,
\end{equation}
where $r = \left| \mathbf{r} \right|$, and $\mathcal{Y}^g_n $ is a spherical surface harmonic of order $n$ (see Appendix A and B for details of the theoretical derivations and expressions). 

At the bubble surface separating the gas and liquid phases, the free slip boundary condition gives $\left( {\bm{\delta}  - \mathbf{nn}} \right) \cdot \bm{\sigma}_l \cdot \mathbf{n} = 0$,  and the normal stress balance gives $\bm{\sigma}_l : \mathbf{nn} - {{\bm{\sigma} _g} : \mathbf{nn}}  = {Ca_S^{-1}}\left( {R_1^{ - 1} + R_2^{ - 1}} \right) $, the normal velocity continuity gives $\mathbf{u}_l \cdot \mathbf{n}$ = $\mathbf{u}_g \cdot \mathbf{n}$. Here, $Ca_S = \mu_l a \dot\gamma / \mathit{\Gamma} $ is the shear capillary number comparing the shear-induced viscosity to capillarity, $R_1$ and $R_2$ are the principle radii of curvature of the interface (Appendix C). The harmonic expressions of the velocities and stresses can be theoretically solved combining with the above boundary conditions, once we know the equation of the bubble surface.

The equation of bubble surface is expressed by considering an additional deformation caused by the rate of change in pressure, $Ca_D f^D$, on a certain initial shape $f^{(0)}$, i.e.,
\begin{equation}    \label{eq_interface}
    r = f^{(0)} + Ca_D f^D ,
\end{equation}
where $f^{(0)}$ is given by Frankel's model \cite{Frankel1970} as $f^{(0)} = 1 + Ca_S \mathcal{F}^S$ (Appendix D), and $Ca_D$ is the dilatation capillary number defined as
\begin{equation}    \label{eq_CaE}
    Ca_D = - \frac{\mu_l a }{\mathit{\Gamma}}  \frac{ \mathrm{d}P}{P \mathrm{d}t} , \quad
\end{equation}
comparing the dilatation-induced viscosity to capillarity, which is positive upon decompressing and negative upon compressing. $f^D$ is also expressed using surface harmonics as $ f^D=\sum {\mathcal{F}^D_n} \left( n \geqslant  0 \right)$ (Appendix D). 

To analytically solve the above equations, we assume the bubble shape deformation and volume dilatation are small ($Ca_S < 1$ and $Ca_D < 1$), thus yielding an influence to the flow belongs to $O$($Ca$). Then the surface harmonics can be solved by the regular perturbation method \cite{Cox1969, Frankel1970}. Results show that only the 0th- and 2nd-order surface harmonics should be considered with the high orders negligible in comparison. Finally, the liquid velocity around the bubble is solved as (Appendix A)
\begin{equation}    \label{eq_finalVelocity_l}
    \mathbf{u}_l = \mathbf{U} + \nabla\left( r^{-1} \mathcal{Y}^l_0 + r^{-3} \mathcal{Y}^l_2 \right) + \frac{1}{2} r^{-3} \mathcal{Z}^l_2 \mathbf{r} , 
\end{equation}
where $\mathcal{Y}^l_n$ is different surface harmonics corresponding to the velocity field of the liquid stokes flow, and $\mathcal{Z}^l_2$ is the second order surface harmonics of the liquid pressure, expressed as ${\mathcal{Z}}^l_2 = 3 \mathbf{Z}^l : \mathbf{rr} r^{-2}$, where
\begin{equation}    \label{zl}
\mathbf{Z}^l = -\frac{2}{3}\mathbf{E} + Ca_S \mathbf{Z}^{l}_S + Ca_D \mathbf{Z}^{l}_D.
\end{equation}
On the right hand of Eq. (\ref{zl}), the first term represents the influence of spherical bubbles with $\mathbf{E} = \frac{1}{2}(\nabla \mathbf{U} + \mathbf{U} \nabla )$ \cite{Taylor1932}, the second term considers the influence of bubble shape deformation under pure shear \cite{Frankel1970}, and the third term evaluates the contribution of bubble volume dilatation with
\begin{equation}    \label{eq_ZE}
    \mathbf{Z}^l_D = -2 \mathcal{F}^D_0 \mathbf{E} - \frac{8}{15} \dot R \mathbf{E},
\end{equation}
where $R = 1 + Ca_D \mathcal{F}^D_0$ is the equivalent radius of the deformed bubble at time $t$, which equals to 1 at $t$ = 0. $\dot R = \mathrm{d}R / \mathrm{d}t$ represents the average expansion speed of the bubble surface. 

Then, the dimensionless tress contribution of the bubble is obtained as Eq. (\ref{eq_finalIntegral}), by substituting the solutions of velocities and stresses into Eq. (\ref{eq_Sigma}).
\begin{equation}    \label{eq_finalIntegral}
    \bm{\Sigma}^g = -\frac{8\pi R^3}{9} (\nabla \cdot \mathbf{u}_g) \bm{\delta} - 4 \pi \bm{\mathcal{Z}}^l ,
\end{equation}
where $\nabla \cdot {\mathbf{u}_g}$ is the average velocity divergence of the gas phase. At a sufficient large pressure that overwhelms the capillary and viscous stresses at the bubble surface, $p_g$ approximately equals $P$ (Appendix E), and thus
\begin{equation}    \label{Divvelocity}
    \nabla \cdot \mathbf{u}_g  =  -\frac{1}{P}\frac{\mathrm{d} P}{\mathrm{d} t} .
\end{equation}
This means that the volumetric dilatation rate \cite{cifuentes2015} of the suspension can be expressed as $\mathcal{\dot{V}} = -\phi \mathrm{d}P/(P\mathrm{d} t) $ where $\phi = \phi_0 [1+t{\mathrm{d}P}/(P\mathrm{d} t)]^{-1}$ is the void fraction of the dilute suspension at time $t$, and the dilatation capillary number can also be expressed as $Ca_D = \mu_l a \mathcal{\dot{V}}/(\mathit{\Gamma} \phi)$.

Substituting Eq. (\ref{eq_finalIntegral}) into Eq. (\ref{eq_integral}), we obtain the constitutive equation of the bubble suspension as
\begin{equation}    \label{eq_totalTau}
    \left\langle \bm{\tau} \right\rangle = 2 \left( 1 + \phi \right) \left\langle \hat{\mathbf{e} } \right\rangle + \left\langle \bm{\tau}_S  \right\rangle + \left\langle \bm{\tau}_D \right\rangle  ,
\end{equation}
where $\hat{\mathbf{e}}$ is the deviatoric rate-of-strain tensor, and $ 1 + \phi $ in the first term on the left hand is the relative viscosity derived by Taylor with spherical bubble assumption \cite{Taylor1932}. $\left\langle \bm{\tau}_S \right\rangle $ evaluates the stress contribution of bubble shape deformation caused by the pure shear flow, given by Frankel \& Acrivos \cite{Frankel1970} as
\begin{equation}
    \left \langle \bm{\tau}_S \right \rangle  = \phi Ca_S \left( - \frac{32}{5}\frac{\mathcal{D}\left \langle \hat{\mathbf{e}} \right \rangle}{\mathcal{D}t} + \frac{48}{35} \mathcal{L}\left[ \left \langle \hat{\mathbf{e}} \right \rangle \cdot \left \langle \hat{\mathbf{e}} \right \rangle  \right] \right),
\end{equation}
where ${\mathcal{D}/\mathcal{D} t}$ is the Jaumann derivative, and $\mathcal{L}[\cdot]$ indicates that only the deviatoric part is retained. 

The last term of Eq. (\ref{eq_totalTau}), $\left\langle \bm{\tau}_D \right\rangle $, represents the stress contribution of the bubble volume dilatation due to the rate of change in pressure, which is expressed as 
\begin{equation}    \label{eq_tauE}
    \left\langle \bm{\tau}_D \right\rangle  = \frac{8}{15} \phi Ca_D \left\langle \hat{\mathbf{e}}\right\rangle ,
\end{equation}
influenced by the volumetric dilatation rate instead of the dilatation amount. Eq. (\ref {eq_totalTau}) indicates that the shear rheology of the dilute bubble suspension is determined by the current void fraction $\phi$, $Ca_S$, and $Ca_D$. When there is no significant pressure variation, $Ca_D \approx 0$, and Eq. (\ref{eq_totalTau}) reduces to Frankel \& Acrivos equation \cite{Frankel1970}, which served as the origin of several followed constitutive equations for bubble suspensions \cite{Llewellin2002,ohie2024}. However, the suspension undergoes dilatation motion upon pressure variation due to the bubble compressibility. Our new constitutive equation demonstrates that the bubble compressibility decreases the viscosity when the pressure is increasing ($Ca_D < 0$), referring to compression-thinning effect, while it increases the viscosity at decreasing pressure ($Ca_D > 0$), referring to decompression-thickening effect. Indeed, the volumetric dilatation rate modifies the suspensions' shear rheological properties, endowed by the intrinsic compressibility of the gas bubbles, which has not been realized in previous studies on shear rheology of bubble suspensions to our best knowledge.

\section{Result and Discussion}\label{sec:3}
\begin{figure}[ht]
\begin{center} 
\includegraphics[width=1\textwidth]{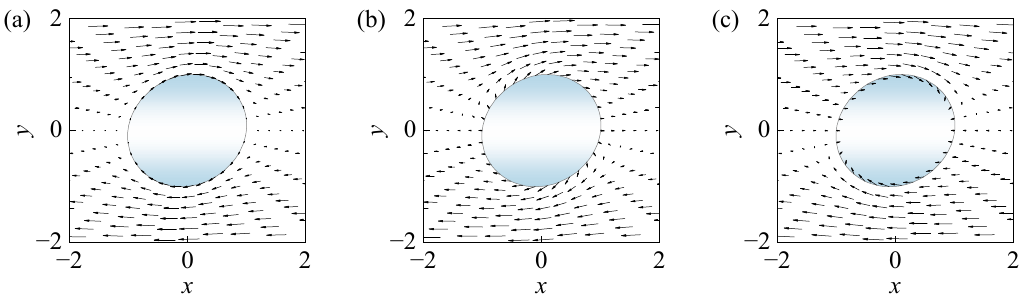}
\caption{\setstretch{1} Velocity field of the liquid surrounding a bubble undergoing a shear flow ($Ca_S = 0.05$) with (a) $Ca_D = 0$ , (b) $Ca_D = 0.05$ (decompressing) and (c) $Ca_D = -0.05$ (compressing). The shade region represents the middle cross section of the bubble at the direction perpendicular to the shear plane. The black arrows represent the liquid velocity vectors.} 
\label{fig:velocity}
\end{center}
\end{figure}

To unveil the behind physics of the compression-thinning behavior caused by bubble compressibility, we analyze the microscopic flow field around a single bubble under a steady shear (Fig. \ref{fig:velocity}). We found that a changing pressure does not apparently affect the bubble shape, but it significantly changes the velocity field around the bubble. The disturbance of the bubble to the liquid flow field only manifests near the bubble, and quickly decays, becoming negligible at a distance of $a$ away from the bubble surface. Without volume dilatation, the liquid flows around the bubble surface, with a zero normal velocity at the interface [Fig. \ref{fig:velocity}(a)]. Thus, the bubble acts like an inviscid droplet, and the macroscopic shear viscosity of the suspension is only influenced by the bubble shape and volume fraction, depending on $Ca_S$ and $\phi_0$ respectively. The presence of bubble leads to the deflection of the liquid streamlines and increases flow resistance, usually resulting a viscosity larger than the liquid viscosity at $Ca_S < 1$ \cite{Taylor1932, Manga1998, Lim2004}. When the pressure is changing, the bubble surface is expanding or shrinking due to the volume dilatation, exerting the liquid near the bubble surface a velocity normal to the interface and allowing the liquid streamlines cross the bubble surface. Specifically, the expanding interface at decreasing pressure forces the external fluid to flow toward the outer layer [Fig.\ref{fig:velocity}(b)]. As a result, the external shear flow is subject to an additional repulsion when approaching the bubbles, which will lead to greater deflection of the liquid streamlines and a stronger flow resistance, thus resulting in a decompression-thickening effect. On the contrary, the shrinking interface at increasing pressure leads to weaker deflection of the liquid streamlines and a smaller flow resistance [Fig.\ref{fig:velocity}(c)], yielding a compression-thinning effect. This compression-thinning behavior is thus determined by the rate of bubble volume dilatation, characterized by the dilatation capillary number $Ca_D$.

We then quantify the contribution of the bubble compressibility to the suspension's shear rheology. With a pressure changing with time, the stress relaxation behavior need to be considered in the shear rheology by applying the operator $\left( 1 + \lambda_t \mathcal{D}/ \mathcal{D}t \right)$ to Eq.(\ref{eq_totalTau}) \cite{Llewellin2002, Mader2013, ohie2024}, where $ \lambda_t = 6 Ca_S / 5 $ is the stress relaxation time. Due to the volume dilatation at a changing pressure, the time derivative of the volume fraction $\mathrm{d} \phi/ \mathrm{d} t = \mathcal{\dot{V}}$ is considered. Then the time evolution of the suspension viscosity can be obtained once we know the $\phi_0$, $Ca_S$, and $Ca_D$ at the initial moment (Appendix F). Referring to the parameter ranges in the displacement or pipeline transport processes of foamy oil reported in previous experiments where significant pressure variations exist \cite{Or2016,Abivin2009,sun2016,Lv2019}, we consider $\mu_l$ = 5 $\mbox{Pa}\cdot \mbox s$, $\mathit{\Gamma}$ = 50 $\mbox{mN}/\mbox{m}$, $ a = 10$-$10^3 \mu \mbox{m}$, $\dot{\gamma}=$ 1-100 s$^{-1}$, $P$ = 1-10 $\mbox{bar}$, $\left| {{\mathrm{d}P} / {\mathrm{d}t}} \right| \leqslant$ 3 ${{\mathrm{bar}} / {\mathrm{s}}}$, which corresponds to a $Ca_S$ = 10$^{-4}$-10 and $Ca_D \sim O(10^{-1})$. To differentiate the viscosity contributions from different mechanisms, we express the relative viscosity of the suspension ($\mu_r = \mu/\mu_l$) as $\mu_r = 1 + \Delta\mu_r^T + \Delta\mu_r^S + \Delta\mu_r^D$, where $\Delta\mu_r^T = \phi$ is the shear-independent viscosity derived by Taylor \cite{Taylor1954}, $\Delta\mu_r^S$ is the pure shear induced viscosity evaluating the real time contribution from a pure steady shear proposed by Frankel \& Acrivos \cite{Frankel1970}, and $\Delta\mu_r^D$ is the dilatation induced viscosity under a changing pressure, as a viscosity correction resulted from the dilatation motion.

\begin{figure}[ht!]
\begin{center} 
\includegraphics[width=0.90\textwidth]{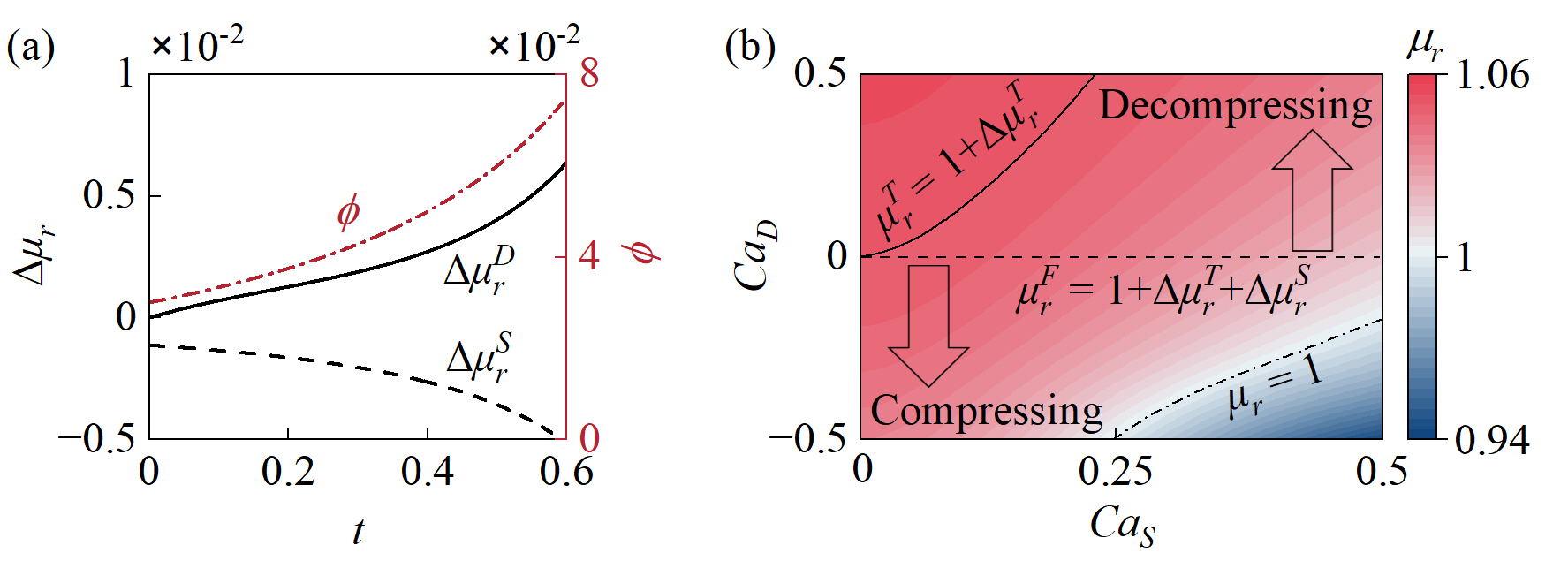}
\caption{\setstretch{1} (a) Time evolution of the relative viscosity $\mu_r$ and void fraction $\phi$ for a bubble suspension of $\phi_0 = 0.03$ at $Ca_S = Ca_D = 0.1$. $\Delta\mu_r^S$ is the pure shear induced viscosity correction \cite{Frankel1970}, and $\Delta\mu_r^D$ is the dilatation induced viscosity correction we proposed. (b) $\mu_r$ of the same suspension at the moment when it expands to $\phi = 0.05$ at different $Ca_S$ and $Ca_D$. $\mu_r^T$ and $\mu_r^F$ are respectively the theoretical predictions by Taylor \cite{Taylor1954} and Frankel \& Acrivos \cite{Frankel1970}.}
\label{fig:shear viscosity}
\end{center}
\end{figure}

Fig. \ref{fig:shear viscosity}(a) shows the time evolution of $\mu_r$ for a suspension of $\phi_0 = 0.03$, suffering from a constant $\dot{\gamma}$ and a negative $\mathrm{d}P/\mathrm{d}t$ with $Ca_S = Ca_D = 0.1$. The void fraction $\phi$ as well as $\Delta\mu_r^T$ increases with time upon decompressing as $\phi = \phi_0 [1+t{\mathrm{d}P}/(P\mathrm{d} t)]^{-1}$. The pure shear induced viscosity $\Delta\mu_r^S$ exhibits a negative correction because the bubble shape deformation facilitates a smaller flow resistance, yielding a shear-thinning effect. In contrast, the dilatation induced viscosity $\Delta\mu_r^D$ shows a positive correction due to the enhanced flow resistance caused by bubble surface expansion. $\Delta\mu_r^D$ and $\Delta\mu_r^S$ both increase with increasing $\phi$ with similar magnitudes, and $\Delta\mu_r^D$ even exceeds $\Delta\mu_r^S$ in the later stage of expansion. The relative viscosity of the same suspension at the moment when it expands to $\phi = 0.05$ with different $Ca_S$ and $Ca_D$ are presented in Fig. \ref{fig:shear viscosity}(b). Generally, $\mu_r$ decreases with $Ca_S$ and increases with $Ca_D$, caused by the different physics behind the viscosity correction induced by pure shear and dilatation. Compared to the viscosity calculated by the equation of Frankel \& Acrivos ($\mu_r^F = 1 + \Delta\mu_r^T + \Delta\mu_r^S$) without considering dilatation effect \cite{Frankel1970}, the $\mu_r$ calculated by our model increases with the decompressing rate and decreases with the compressing rate, highlighting the compression-thinning or decompression-thickening behavior stemmed from the dilatation motion of the suspension endowed by bubble compressibility. At a large $\left|Ca_D\right|$, this dilatation induced rheological effect can play a leading role in the shear viscosity of the suspension, yielding a $\mu_r$ larger than Taylor's prediction ($1+\phi$) at a large decompressing rate and a $\mu_r$ smaller than 1 at a large compressing rate. 

It seems unexpected that the bubble compressibility can profoundly modify the shear viscosity of the suspension, considering the minor volume dilatation of the dilute suspension of small $\phi$. Using the parameter values mentioned before, a $Ca_D$ of 0.1 corresponds to a volumetric dilatation rate of $\mathcal{\dot{V}} =$ 3\% s$^{-1}$ for $a$ = 1 mm at 1 atm, which means that the volume variation of the bubble suspension is usually unobtrusive at high frequency pressure fluctuations in industry \cite{wang2012,ubertini2006}. Exemplified with the significant compression-thinning effect for dilute suspensions upon pressure variation, we highlight the critical role of microscopic flows around the single bubbles in shaping the macroscopic properties of the bubble suspensions. We envision that the shear viscosity contributed by the dilatation motion may be more important for dense bubble suspensions, which remains to be explored in future studies. In addition, we note that the existing rheology experiments are usually conducted at a simple shear flow without violent pressure changing. It is thus interesting to explore the implications of our model by rheological measurements with significant rate of change in pressure regarding the complex flows in practical situations.

\section{Conclusion}\label{sec:4}
In summary, we theoretically investigate the shear rheology of dilute bubble suspensions in consideration of the volume dilatation endowed by bubble compressibility. Our constitutive equation reports a compression-thinning behavior for the first time, manifesting a decreasing shear viscosity upon compressing and an increasing shear viscosity upon decompressing. This peculiar behavior is microscopically derived from the fact that the shrinking motion of the bubble surface upon compressing weakens the flow resistance of the external liquid under shear, and vice versa. We show the dilatation induced viscosity can exceed the shear induced viscosity and dominate the suspension's shear viscosity in practical scenarios. Our study not only advances the understanding on the rheological behaviors of dispersion systems, but also may provide guidance for the flow control of bubble suspensions in practice.

\bigskip

\section*{Appendix A: Solution of Stokes liquid flow by spherical harmonic analysis} \label{appA}
According to the standard solution of Stokes Equations given by \citet{Lamb1932}, the local velocity of the liquid phase around the bubble, $\mathbf{u}_l$, is expressed as
\begin{eqnarray}    \label{eq_ul}
    {\mathbf{u}}_l & = & \mathbf{U} + \sum\limits_{n = 0}^{ + \infty } { \left[ {\nabla \left(r^{-n-1}{{\mathcal{X}}^l_{n}}\right) \times \mathbf{r} + \nabla \left(r^{-n-1}{{\mathcal{Y}}^l_{n}}\right) } \right. } \nonumber\\
    && \quad\quad\quad \left. - \frac{{n - 2}}{{2n\left( {2n - 1} \right)}}{r^2}\nabla \left(r^{-n-1}{{\mathcal{Z}}^l_{n}} \right) + \frac{{n + 1}}{{n\left( {2n - 1} \right)}}\mathbf{r} \left(r^{-n-1}{{\mathcal{Z}}^l_{n}} \right) \right],
\end{eqnarray}
and the local liquid pressure, $p_l$, is derived as
\begin{equation}    \label{eq_pl}
    p_l = P + \sum\limits_{n = 1}^{ + \infty } { \left(r^{-n-1}{{\mathcal{Z}}^l_{n}} \right) }.
\end{equation}
where all parameters are dimensionless, $\mathcal{X}^l_n$, $\mathcal{Y}^l_n$ and $\mathcal{Z}^l_n$ are three different spherical surface harmonics of order $n$. According to the orthogonality relations, these three kinds of harmonics can be solved analytically using the regular perturbation method \citep{Cox1969, Frankel1970}, e.g. $ \mathbf{u} = \bar{\mathbf{u}} + Ca_S \tilde{\mathbf{u}}_S + Ca_D \tilde{\mathbf{u}}_D$ where the superscripts $\bar \cdot$ and $\tilde \cdot$ represent the $O(1)$ and $O(Ca)$ components, respectively.

Firstly, the zeroth order harmonics should satisfy the following relationships:
\begin{equation}    \label{eq_y0l}
    \frac{\mathcal{Y} _{0}^{l}}{R^3} 
    = - \frac{1}{R}\frac{\mathrm{d}R}{\mathrm{d}t} 
    = - \frac{1}{3} \left( \nabla \cdot \mathbf{u}_g \right),
\end{equation}
\begin{equation}    \label{eq_final_pg}
    p_g = P\left( t \right) +2Ca_{S}^{-1}\frac{1}{R} + \frac{4}{R}\frac{\mathrm{d}R}{\mathrm{d}t} ,
\end{equation}
where the equivalent bubble radius $R$ is related to the 0th-order harmonic $\mathcal{F}_0^D$. According to the ideal gas law and mass conservation, $R$ is determined by the pressure inside the bubble, which means that $p_g R^3$ is always equal to the initial value $\left. p_g \right|_{t=0}$. Hence, $\mathcal{Y} _{0}^{l}$ and $\nabla \cdot \mathbf{u}_g$, the volumetric dilatation rate of the single bubble, can be obtained accurately by solving the set of differential equations, Eqs.~(\ref{eq_y0l}) and (\ref{eq_final_pg}). At a sufficient large pressure that overwhelms the capillary and viscous stresses at the bubble surface, $p_g$ approximately equals $P$ (Appendix E), we have $\nabla \cdot \mathbf{u}_g = - {\mathrm{d}P} / (P\mathrm{d} t)$.

Then the coefficients of the second order harmonics can be derived as
\begin{equation}
    {\mathbf{Y}}^{l} = 
    Ca_S \left( \frac{2}{15} \frac{\mathcal{D} \mathbf{E}}{\mathcal{D}t} - \frac{2}{45} \mathcal{L}\left[ \mathbf{E} \cdot \mathbf{E} \right] \right)
    + Ca_D \left( - \frac{8}{45} \mathbf{E} \right) ,
\end{equation}
\begin{equation}
    {\mathbf{Z}}^{l} = -\frac{2}{3} \mathbf{E}
    + Ca_S \left( \frac{32}{15} \frac{\mathcal{D} \mathbf{E}}{\mathcal{D}t} - \frac{48}{105} \mathcal{L}\left[ \mathbf{E} \cdot \mathbf{E} \right] \right)
    + Ca_D \left( - 2 {\mathcal{F}}_0^D \mathbf{E} - \frac{8}{45} \mathbf{E} \right),
\end{equation}
and the second order harmonics is obtained by ${\mathcal{Y}}_2^{l} = 3 {\mathbf{Y}}^{l} \pmb{:} \mathbf{rr} r^{-2} $ and ${\mathcal{Z}}_2^{l} = 3 {\mathbf{Z}}^{l} \pmb{:} \mathbf{rr} r^{-2} $. The odd order harmonics are zero and the harmonics with an order larger than 2 is negligible compared to the zeroth and second orders.

\section*{Appendix B: Solution of the gas velocity inside the bubble by spherical harmonic analysis}\label{appB}
The velocity potential ${\mathit{\Phi} _g}$ for the inviscid gas is used to describe the velocity inside the bubble, that is $\nabla {\mathit{\Phi} _g} = {\mathbf{u}_g}$. Hence, 
\begin{equation}\label{eq2.7}
    {\nabla ^2}{\mathit{\Phi} _g} =  - \frac{1}{{{p_g}}}\frac{{\mathrm{d} {p_g}}}{{\mathrm{d} t}} .
\end{equation}

Assume that the solution of this non-homogeneous differential equation [Eq.~(\ref{eq2.7})] can be expressed as the sum of a general solution ${\theta _g}$ and a special solution ${\vartheta _g}$, yielding
\begin{equation}\label{eq2.8}
    {\mathit{\Phi} _g} = {\theta _g} + {\vartheta _g},
\end{equation}
where ${\nabla ^2}{\theta _g} = 0$ and ${\nabla ^2}{\vartheta _g} = -\mathrm{d}p_g / p_g \mathrm{d}t$. According to the solution of Laplace Equation, the general solution ${\theta _g}$ can be directly expressed by the spherical surface harmonics, i.e. $\theta _n^g = {r^n} \mathcal{Y} _n^g$. Meanwhile, since the pressure inside the bubble is independent of the spatial position, as given by $\nabla p_g = 0$, the special solution ${\vartheta _g}$ can be written as $- r^2 \mathrm{d} p_g / 6 p_g \mathrm{d}t$. 

Hence, the velocity inside the bubble is finally expressed as
\begin{equation}\label{eq_ub}
    {\mathbf{u}_g} = - \frac{1}{{6{p_g}}}\frac{{\mathrm{d}{p_g}}}{{\mathrm{d}t}}\nabla {r^2} + \sum\limits_{n = 0}^{ + \infty } {\nabla \left( {r^n} \mathcal{Y} _n^g \right)}.
\end{equation}

Through the perturbation method, we obtain $\mathcal{Y}_n^g = 0$ for $n \ne 2$, and ${\mathcal{Y}}_2^{g} = 3 {\mathbf{Y}}^{g} \pmb{:} \mathbf{rr} r^{-2} $, where
\begin{equation}
    {\mathbf{Y}}^g = Ca_D \left( -\frac{1}{9} \mathbf{E} \right).
\end{equation}

\section*{Appendix C: Sum of the principal curvatures of the bubble surface}\label{appC}
According to the tensor analysis method \citep{mcconnell2014}, the sum of the principle curvatures can be expressed as
\begin{equation} \label{eqA9}
    R_1^{-1} + R_2^{-1} = 2 -2 \Sigma Ca f - \Sigma Ca f_{,kk} +  2 \left( \Sigma Ca f \right)^2 + 2 \left( \Sigma Ca f \right) \left( \Sigma Ca f_{,kk} \right) + O ( Ca ^3 ) ,
\end{equation}
where $\Sigma Caf$ represents any small deformation of the bubble surface based on the sphere ($r=1$), and the subscripts comma and $kk$ respectively denote the derivation operator and Einstein summation convention in tensor analysis. 

The above equation can be rewritten in the form of spherical surface harmonics, yielding
\begin{eqnarray}
    R_1^{-1}+R_2^{-1} & = & 2 + 4 Ca_S \mathcal{F}^S + Ca_D \sum_{n=0}^{+\infty}{\left( n^2+n-2 \right) \mathcal{F}_{n}^{D}} \nonumber\\
    && - 10 Ca_{S}^{2} \left( \mathcal{F}^{S} \right)^2 - 2 Ca_S Ca_D \mathcal{F}^S \sum_{n=0}^{+\infty}{\left( n^2+n+4 \right) \mathcal{F}_{n}^{D}} \nonumber\\
    && - 2 Ca_{D}^{2} \sum_{n=0}^{+\infty}{\mathcal{F}_{n}^{D}} \sum_{n=0}^{+\infty}{\left( n^2-n-1 \right) \mathcal{F}_{n}^{D}}  + O\left( Ca^3 \right), 
\end{eqnarray}
where $\mathcal{F}_{n,kk} = -n(n+1)\mathcal{F}_n$ can be derived from the Laplace Equation at $r=1$ \citep{Cox1969}.

\section*{Appendix D: Equation of bubble surface by spherical harmonics analysis}\label{appD}
In Frankel's model \citep{Frankel1970}, the bubble shape under pure shear can be described as $r = 1 + Ca_S \mathcal{F}^S = 1 + 3 Ca_S {\mathbf{F}}^S\pmb{:}\mathbf{rr} r^{-2}$, where
\begin{equation}    \label{eq_shape_Frankel}
    {\mathbf{F}}^S + \frac{6}{5} Ca_S \frac{ \mathcal{D} {\mathbf{F}}^S }{\mathcal{D} t} = \frac{2}{3} \mathbf{E} + \frac{128}{105} Ca_S \mathcal{L} \left[ \mathbf{E} \cdot \mathbf{E} \right] .
\end{equation}

Considering a macroscopic pressure $P$ changing in time, the additional deformation $Ca_D f^D$ due to the bubble compressibility is expressed as the sum of the 0th- and 2nd-order surface harmonics, $Ca_D \mathcal{F}_0^D$ and $Ca_D \mathcal{F}_2^D$, which respectively represent the volume dilatation and additional shape deformation of the bubble. For the former, $\mathcal{F}_0^D$ can be linked to the equivalent radius $R$, given by $R = \sqrt[3]{ {\left. P \right|_{t=0}}/{P} }$. For the latter, the coefficient tensor of $\mathcal{F}_2^D$ is determined by the following differential equation:
\begin{equation} \label{eq_shape_add}
    {\mathbf{F}}^D + \frac{6}{5}Ca_S\frac{\partial {\mathbf{F}}^D}{\partial t} = \frac{4}{3} Ca_S  {\mathcal{F}}_0^D \mathbf{E} - \frac{4}{15} Ca_S \mathbf{E}.
\end{equation}

From Eqs.~(\ref{eq_shape_Frankel}) and (\ref{eq_shape_add}), we can find that the leading factor is determined by the $O$(1) component in Eq.~(\ref{eq_shape_Frankel}), that is $2\mathbf{E}/3$, and the relaxation time $\lambda_t$ is only related to the shear capillary number, i.e. $\lambda_t = 6Ca_S/5$. Hence, the bubble shape is still led by the shear effect in the flow field (Figure~\ref{fig:velocity} in the main text).

For simplicity, these differential equations can be simplified by the successive substitution method \citep{Frankel1970}, as the magnitude of the time derivative terms in Eqs.~(\ref{eq_shape_Frankel}) and (\ref{eq_shape_add}) are usually less than unity. Thus
\begin{equation}
    {\mathbf{F}}^S = \frac{2}{3} \mathbf{E} + Ca_S \left( - \frac{4}{5}  \frac{\mathcal{D} \mathbf{E}}{\mathcal{D} t} + \frac{128}{105} \mathcal{L} \left[ \mathbf{E} \cdot \mathbf{E} \right] \right) + O(Ca^2).
\end{equation}
\begin{equation} \label{eq_F2D_2}
    {\mathbf{F}}^D = \frac{4}{3} Ca_S  {\mathcal{F}}_0^D \mathbf{E} - \frac{4}{15} Ca_S \mathbf{E} + O(Ca^2).
\end{equation}

Finally, the equation of the bubble surface can be expressed as 
\begin{equation}
    r = 1 + Ca_S \left( 3 {\mathbf{F}}^S \pmb{:} \mathbf{rr} r^{-2} \right) 
    + Ca_D \left( \mathcal{F}^D_0 + 3 {\mathbf{F}}^D \pmb{:} \mathbf{rr} r^{-2} \right).
\end{equation}

\section*{Appendix E: Gas pressure in bubble based on sufficient macroscopic pressure assumption}\label{appE}
Through dimensional analysis, the magnitude of the interface tension may not exceed the order of $10^{4}$ N/m$^2$ when the bubble size is larger than $10$ $\mu$m. Meanwhile, the $O$(1) viscous stress $4 {\mathrm{d}R} / {R \mathrm{d}t}$ in Eq.~(\ref{eq_final_pg}) is always less than the $O(Ca^{-1}_S)$ interface tension due to the small deformation hypothesis. In these situations, even at a macroscopic pressure $P$ as low as 1 atm ($10^5$ Pa), it is much lager than other terms on the right hand of the stress balance equation [Eq.~(\ref{eq_final_pg})]. Hence, the pressure inside the bubble, $p_g$, can approximately equal to $P$, and 
\begin{equation}    \label{eq_div_ug and R}
    \nabla \cdot \mathbf{u}_g = -\frac{1}{P}\frac{\mathrm{d} P}{\mathrm{d} t}.
\end{equation}

\section*{Appendix F: Shear viscosity of bubble suspensions considering the stress relaxation}\label{appF}
For Frankel's model \citep{Frankel1970} without considering the effect from dilatation motion, the constitutive equation with stress relaxation is expressed as 
\begin{equation}    \label{eq_constitutiveEqu_frankel}
    \left \langle \bm{\tau} \right \rangle + \frac{6}{5} Ca_S \frac{\mathcal{D} \left \langle \bm{\tau} \right \rangle }{\mathcal{D}t} = 2\left( 1 + \phi_0 \right) \left \langle \hat{\mathbf{e}} \right \rangle +  Ca_S \left[ \frac{12}{5} \left( 1 - \frac{5}{3} \phi_0 \right) \frac{\mathcal{D} \left \langle \hat{\mathbf{e}} \right \rangle }{\mathcal{D}t} + \frac{48}{35} \phi_0 \mathcal{L}\left[ \left \langle \hat{\mathbf{e}} \right \rangle \cdot \left \langle \hat{\mathbf{e}} \right \rangle \right] \right],
\end{equation}

Under a steady simple shear flow that $\left \langle {\mathbf u} \right \rangle = (y, 0, 0)$, the relative shear viscosity given by Frankel's model [Eq.~(\ref{eq_constitutiveEqu_frankel})] can be obtained that
\begin{equation}
    \mu_r^F = 1 + \phi_0 \frac{1 - \frac{12}{5}Ca_S^2}{1 + \left( \frac{6}{5} Ca_S \right) ^2}
\end{equation}

To evaluate the real time contribution of the shear effect considered by \citet{Frankel1970}, the initial void fraction $\phi_0$ and the shear capillary number are replaced by $\phi$ and $Ca_S R$, respectively. Considering the viscosity of suspensions with spherical bubbles derived by \citet{Taylor1932}, the pure shear induced viscosity $\Delta \mu^S_r$ can be expressed as
\begin{equation}
    \Delta \mu^S_r = - \phi \frac{96 \left( Ca_S R \right) ^2}{25 + 36 \left( Ca_S R \right) ^2} 
\end{equation}

With the consideration of stress relaxation, the constitutive equation of the bubble suspension we proposed [Eq.~(\ref{eq_totalTau}) in the main text] can be updated to 
\begin{eqnarray}    \label{eq_constitutiveEqu_now}
    \left \langle \bm{\tau} \right \rangle + \frac{6}{5} Ca_S \frac{\mathcal{D} \left \langle \bm{\tau} \right \rangle }{\mathcal{D}t} & = & 2\left( 1 + \phi \right) \left \langle \hat{\mathbf{e}} \right \rangle + \frac{44}{15} \phi Ca_D \left \langle \hat{\mathbf{e}} \right \rangle
    \nonumber\\
    &&+  Ca_S \left[ \frac{12}{5} \left( 1 - \frac{5}{3} \phi \right) \frac{\mathcal{D} \left \langle \hat{\mathbf{e}} \right \rangle }{\mathcal{D}t} + \frac{48}{35} \phi \mathcal{L}\left[ \left \langle \hat{\mathbf{e}} \right \rangle \cdot \left \langle \hat{\mathbf{e}} \right \rangle \right] \right],
\end{eqnarray}
where the rate of change of void fraction $\phi$ is related to the volumetric dilatation rate $\dot{\mathcal{V} }$, because
\begin{equation}    \label{eq_phi & dphi}
    \phi = \frac{\frac{4}{3}\pi R^3}{ \frac{4}{3}\pi R^3 + V_l },
    \quad \mbox{and} \quad
    \frac{\mathrm{d}\phi}{\mathrm{d}t} = \phi \frac{3}{R} \frac{\mathrm{d}R}{\mathrm{d}t} + O(\phi^2) = - \phi \frac{1}{P} \frac{\mathrm{d}P} {\mathrm{d}t} + O(\phi^2),
\end{equation}
where $V_l$ is the liquid volume which is assumed to be a constant, and the $O(\phi^2)$ terms are neglected for the dilute system. Based on Eq.~(\ref{eq_phi & dphi}), $Ca_S \mathrm{d} \phi / \mathrm{d}t=\phi Ca_D$.
To obtain the shear stress related to the viscosity, we should solve the following differential equations for the components of the deviatoric stress $\left \langle \bm{\tau} \right\rangle$:
\begin{equation}\label{eq5.1}
    \left( {1 + \frac{6}{5}Ca_S\frac{\partial }{{\partial t}}} \right){\tau _{11}} - \frac{6}{5}Ca_S{\tau _{12}} = \frac{6}{5}\left( { - 1 + \frac{{37}}{{21}}\phi } \right)Ca_S,
\end{equation}
\begin{equation}\label{eq5.2}
    \left( {1 + \frac{6}{5}Ca_S\frac{\partial }{{\partial t}}} \right){\tau _{22}} + \frac{6}{5}Ca_S{\tau _{12}} = \frac{6}{5}\left( {1 - \frac{{11}}{7}\phi } \right)Ca_S,
\end{equation}
\begin{equation}\label{eq5.3}
    \left( {1 + \frac{6}{5}Ca_S\frac{\partial }{{\partial t}}} \right){\tau _{33}} =  - \frac{8}{{35}}\phi Ca_S,
\end{equation}
\begin{equation}\label{eq5.4}
    \left( {1 + \frac{6}{5}Ca_S\frac{\partial }{{\partial t}}} \right){\tau _{12}} + \frac{3}{5}Ca_S\left( {{\tau _{11}} - {\tau _{22}}} \right) = 1 + \phi  + \frac{{22}}{{15}}\phi Ca_D.
\end{equation}
where the time derivatives of the stress components $\tau_{ij}$ need to be taken into account due to the pressure changing with time. The analytical solutions of Eq.~(\ref{eq_constitutiveEqu_frankel}) corresponding to the pure shear flow \citep{Pal2003} can be selected as the initial value $(t=0)$ of the set of differential equations of (\ref{eq5.1})-(\ref{eq5.4}). Then we can obtain the value of the relative shear viscosity $\mu_r$, which is equal to the value of the dimensionless shear stress $\tau_{12}$. Finally, the dilatation induced viscosity $\Delta \mu _r^D$ can be calculated as $\mu_r - 1 - \Delta \mu^T_r - \Delta \mu^S_r$.

\bigskip
\bibliography{References}

\end{document}